\documentclass[twocolumn]{article}  
\usepackage{natbib}
\usepackage{geometry}
\usepackage{graphicx}

\usepackage{subcaption}
\usepackage{amsmath}
\usepackage{amssymb}
\usepackage{amsfonts}
\usepackage{amsthm}
\usepackage{bm}
\usepackage{enumitem}

\usepackage{times}
\usepackage{soul}
\usepackage{url}
\usepackage[hidelinks]{hyperref}
\usepackage[utf8]{inputenc}
\usepackage{caption}
\usepackage{graphicx}
\usepackage{amsthm}
\usepackage{booktabs}
\usepackage{algorithm}
\usepackage{algorithmic}
\usepackage{diagbox}

\usepackage{flushend}

\makeatletter
\newcommand{\printfnsymbol}[1]{%
  \textsuperscript{\@fnsymbol{#1}}%
}
\makeatother

\begin{document}

\title{Networked Multi-Agent Reinforcement Learning with Emergent Communication}  

\author{\textbf{Shubham Gupta}\thanks{Equal Contribution} and \textbf{Rishi Hazra}\printfnsymbol{1} and \textbf{Ambedkar Dukkipati}\\
Department of Computer Science and Automation\\
Indian Institute of Science\\
Bangalore - 560012, India.\\
\texttt{[shubhamg, rishihazra, ambedkar]@iisc.ac.in}}

\date{}

\maketitle

\begin{abstract} Multi-Agent Reinforcement Learning (MARL) methods find optimal policies for agents that operate in the presence of other learning agents. Central to achieving this is how the agents coordinate. One way to coordinate is by learning to communicate with each other. Can the agents develop a language while learning to perform a common task? In this paper, we formulate and study a MARL problem where cooperative agents are connected to each other via a fixed underlying network. These agents can communicate along the edges of this network by exchanging discrete symbols. However, the semantics of these symbols are not predefined and, during training, the agents are required to develop a language that helps them in accomplishing their goals. We propose a method for training these agents using emergent communication. We demonstrate the applicability of the proposed framework by applying it to the problem of managing traffic controllers, where we achieve state-of-the-art performance as compared to a number of strong baselines. More importantly, we perform a detailed analysis of the emergent communication to show, for instance, that the developed language is grounded and demonstrate its relationship with the underlying network topology. To the best of our knowledge, this is the only work that performs an in depth analysis of emergent communication in a networked MARL setting while being applicable to a broad class of problems\footnote{short paper accepted at AAMAS 2020}.
\end{abstract}


\section{Introduction}
\label{section:introduction}
Co-existing intelligent agents affect each other in non-trivial ways. Any change in one agent's policy modifies the other agent's perception about the environment dynamics. This turns the environment non-stationary and the learning problem becomes hard. Approaches that try to independently learn optimal behavior for agents do not perform well in practice~\citep{Tan:1993:MultiAgentReinforcementLearningIndependentVsCooperativeAgents}. Thus, it is important to develop models that have been tailored towards training multiple agents simultaneously. Multi-agent reinforcement learning (MARL) provides the formalism to do so.

In this paper, we consider a multi-agent setting where a certain number of cooperative agents co-exist in an environment that can only be partially observed by each of them. Further, we assume that these agents are interconnected via a fixed network topology and that they have been endowed with the ability to communicate with each other along the edges of this network. The objective of agents is to learn a communication protocol so as to cooperatively maximize the rewards provided to them by the environment. This problem setting has two interesting features:

\noindent
\textbf{1. Communication:} Communication allows the agents to augment their local observations with additional information that is necessary to achieve global cooperation. It also enables a more dynamic form of coordination among agents (for example, agents soliciting help from their peers after entering a particular state) as opposed to the well known centralized training, decentralized execution paradigm \citep{NIPS2017_7217} where the agents are trained together but must act independently in a decentralized fashion post deployment.

\noindent
\textbf{2. Network of agents:} We assume that agents can only communicate along the edges of a fixed underlying network. We believe that this is a more practical scenario as direct communication between all agents may not always be possible due to constraints like geographical separation, limited communication bandwidth and so on. Moreover, having a fixed network allows us to study the relationship between the emergent communication and the topology of the underlying network (Section \ref{section:experiments}). We restrict our attention to discrete communication, i.e., agents communicate by exchanging discrete symbols. This has been done to: \textbf{(i)} facilitate analysis of emergent language; and \textbf{(ii)} conserve communication bandwidth which is important for practical applications.

Emergent communication has been studied both in the context of referential games \citep{NIPS2017_6810,Mordatch2017EmergenceOG} and for developing abstract strategies \citep{CaoEtAl:2018:EmergentCommunicationThroughNegotiation,GuptaEtAl:2019:OnVotingStrategiesAndEmergentCommunication}. While these approaches offer many insights into emergent communication, their applicability is limited in practice. More generic approaches, like TarMAC \citep{DasEtAl:2019:TarMACTargetedMultiAgentCommunication} which explores the utility of attention in deciding whether a given pair of agents may communicate with each other or not, have been proposed but they lack in-depth analysis of communication. As opposed to \citep{DasEtAl:2019:TarMACTargetedMultiAgentCommunication}: \textbf{(i)} we use discrete communication which helps us in performing a detailed analysis of communication and, \textbf{(ii)} we only allow agents to communicate via an underlying network, i.e., certain agents may never directly communicate with each other, thus the agents must learn to perform well in a more practical but also more constrained environment. This also enables the study of the effect of underlying network topology on emergent language.

Many real world problems can be cast in this framework. For example, consider the problem of intelligently managing traffic in a city. The nodes in the network (i.e., the agents) correspond to traffic controllers and the edges correspond to roads. The controllers must act cooperatively to ensure a smooth flow of traffic by maximizing an appropriate notion of reward. We present the traffic management problem (Section \ref{section:intelligent_traffic_management}) as a particular instantiation of the proposed abstract MARL problem (Section \ref{section:problem_setup}). Although we provide a number of other concrete examples in Section \ref{section:problem_setup}, for clarity of exposition and to be more concrete, in this paper, we only focus on the problem of intelligently managing traffic as a case study. This problem was chosen because of the easy availability of high quality simulators and because it subsumes a number of interesting problems like routing of network packets, air traffic control and so on.

Our main contributions are: formulation of the abstract MARL problem with networked agents and emergent communication as stated above (Section \ref{section:problem_setup}), demonstration of the effectiveness of proposed approach using traffic management as a case study (Section \ref{section:intelligent_traffic_management}) and most importantly analysis of the emergent communication (Section \ref{section:experiments}) to investigate: \textbf{(i)} utility of communication; \textbf{(ii)} grounding of language (i.e., whether \textit{words} in the language refer to physical actions); and \textbf{(iii)} interplay between the underlying network topology and emergent language. 


\section{Related Work}
\label{section:related_work}

Independent training of agents in a MARL setting often results in poor performance \citep{Tan:1993:MultiAgentReinforcementLearningIndependentVsCooperativeAgents}. A straightforward way to address this issue is to view the collection of agents as a single meta-agent and then train this meta-agent using existing single agent reinforcement learning techniques. However, such an approach is not scalable as the action space of the meta-agent grows exponentially with the number of agents.

The centralized training, decentralized execution paradigm \citep{NIPS2017_7217} avoids non-stationarity issues during training by providing the global state information to all agents. Once trained, these agents can be executed independently of each other. There are two common issues associated with this paradigm: \textbf{(i)} as the number of agents grows, the centralized training step gets harder; and \textbf{(ii)} as during the test time the agents have to act independently, this strategy is not optimal in scenarios where a more dynamic form of context dependent coordination is required. Decentralized training, decentralized execution approaches \citep{WenEtAl:2019:ProbabilisticRecursiveReasoningForMultiAgentReinforcementLearning} address the first issue but not the second one.

Communication enables agents to exchange information even during the deployment phase, thus providing a solution to the problem of achieving dynamic coordination. Certain approaches use a fixed communication protocol \citep{ZhangEtAl:2018:FullyDecentralizedMultiAgentReinforcementLearningWithNetworkedAgents} while others learn to communicate while solving the desired task \citep{NIPS2016_6398,Mordatch2017EmergenceOG,CaoEtAl:2018:EmergentCommunicationThroughNegotiation,GuptaEtAl:2019:OnVotingStrategiesAndEmergentCommunication,DasEtAl:2019:TarMACTargetedMultiAgentCommunication}. Some of these approaches use continuous communication \citep{NIPS2016_6398,DasEtAl:2019:TarMACTargetedMultiAgentCommunication} while others use discrete communication. We use emergent discrete communication for reasons described in Section \ref{section:introduction}. As opposed to existing methods, the emergent communication in our method is influenced by the structure of the underlying network.

MARL over networks has also been studied in \citep{ZhangEtAl:2018:FullyDecentralizedMultiAgentReinforcementLearningWithNetworkedAgents} in a fully decentralized training setup where local parameters of the agents are shared through communication. However, \citep{ZhangEtAl:2018:FullyDecentralizedMultiAgentReinforcementLearningWithNetworkedAgents} assume that all agents have full access to the global environment state while we allow partial observability. \citep{NIPS2016_6398} also use communication in traffic networks, but: \textbf{(i)} we formulate this as a network problem with traffic controllers as agents as opposed to treating vehicles as agents; and \textbf{(ii)} we use network restricted, discrete communication. Our approach is very similar to DIAL~\citep{10.5555/3157096.3157336} which uses a centralized training and a decentralized execution with communication channels. However, we use an attention mechanism in our setup to prioritize incoming messages in addition to the use of Gumbel-Softmax~\citep{journals/corr/JangGP16} which yields better quality gradients during training.

The simplest and one of the most widely used approaches to tackle the traffic management problem is the Fixed-time Control~\citep{Miller1963SettingsFF,Webster1958}, which uses a predefined cycle for planning. Self-Organizing Traffic Light Control (SOTL)~\citep{SOTL} method switches the traffic lights when the number of vehicles crosses a predefined threshold. These conventional traffic control methods rely on assumptions that do not hold in practice. In the recent past, reinforcement learning methods have also been used for dynamically adapting to the traffic conditions~\citep{6958095,7508798,Wei:2018}. However, these approaches consider the agents to be independent of one another, thus leading to a non-stationary environment. 


\section{Proposed Markov Games with Emergent Communication}
\label{section:problem_setup}

Markov games \citep{Littman:1994:MarkovGamesAsAFormulationForMARL} are used for modeling multi-agent environments. Let $\mathcal{S}$  be the set of all environment states and $\mathbf{s}^{(t)} \in \mathcal{S}$ be the state of environment at time $t$. At time $t$, observation function $f_i:\mathcal{S} \rightarrow \mathcal{O}_i$ yields the observation, $\mathbf{o}_i^{(t)} \in \mathcal{O}_i$, for agent $i$ where $\mathcal{O}_i$ is its observation space. Based on the observation, each agent $i$ chooses an action $\mathbf{a}_i^{(t)} \in \mathcal{A}_i$ using its policy $\pi_i$, i.e., $\mathbf{a}_i^{(t)} \sim \pi_i(\cdot \vert \mathbf{o}_i^{(t)})$. The state is then updated using the transition function $\mathcal{T}$: $\mathcal{S} \times \mathcal{A}_1 \times \dots \times \mathcal{A}_N \times \mathcal{S} \rightarrow [0, 1]$. Agents receive rewards via the reward functions $r_i$: $\mathcal{S} \times \mathcal{A}_1 \times \dots \times \mathcal{A}_N \rightarrow \mathbb{R}$, $i = 1, 2, \dots, N$. The goal is to find optimal policies $\pi^*_{i}$ : $\mathcal{O}_i \times \mathcal{A}_i \rightarrow [0, 1]$, $i = 1, 2, \dots, N$ that maximize the expected long term reward, $\mathcal{R}_i = \mathrm{E}_{\bm{\pi}} [\sum_{t} \gamma^t r_{i}^{(t)}]$ for each agent $i$. Here $r_i^{(t)}$ is agent $i$'s reward at time $t$, $\gamma \in (0, 1]$ is the discount factor and, $\bm{\pi} = \pi_1 \times \pi_2 \times \dots \times \pi_N$.

We model the problem as a Markov game with two additional assumptions: \textbf{(i)} let $\mathcal{V} = \{1, 2, \dots, N\}$ be the set of agents, we assume that these agents are interconnected via an underlying network with edge set $\mathcal{E}$; and \textbf{(ii)} agents can communicate with their immediate neighbors in the underlying network using messages drawn from a discrete space. A message broadcasted by an agent at time $t$ is received by its neighbors at time $t+1$. The observation space of agents is augmented to also consider the messages received by it.

While agents maximize their own rewards, during training, we allow them to provide feedback on the quality of received messages to their neighbors via gradients. This provides incentives to the agents to send messages that help their neighbors in maximizing their rewards thereby ensuring that the agents act in a cooperative manner. Communication between competitive agents is an interesting problem but it poses a different set of challenges (for example, agents may mislead each other by conveying false information), hence, in order to perform a meaningful analysis of emergent communication, we restrict ourselves to a cooperative setting in this paper.

There are numerous practical scenarios that can be modeled this way. Consider, for instance, an intelligent electrical distribution network: in this application, agents represent power stations and the underlying network corresponds to the electrical grid. Each agent has a production capacity and it may choose to share the power generated by it with a neighboring agent (action space). All agents observe various attributes like power requirements, consumer demand and so on (observation space) and they have to meet the local demand which changes stochastically. Rewards $r_i$ measure the success of agents in meeting their local demands without wasting surplus power. As another application, consider a supply chain where interconnected warehouses (agents) have to manage their inventory to meet the local demand. As before, agents can choose to ship goods that they have in their inventory to their neighbors (action space). The warehouses have to meet stochastically changing demands and must learn to communicate effectively in order to share goods so that an appropriate level of inventory is maintained at each warehouse.

As mentioned earlier, in this paper, we study intelligent traffic management as a concrete instantiation of the proposed MARL problem (Section \ref{section:intelligent_traffic_management}). For all real world applications that we have mentioned above, transparency in decision making is important and using discrete communication is a step in this direction as we believe that it is more interpretable. Towards this end, we demonstrate that the emergent language is grounded, i.e. messages correspond to actions in Section \ref{section:experiments}. If agents communicate using discrete symbols, humans can potentially inspect and interpret the conversation logs.


\subsection{Learning Policies with Communication}
\label{section:proposed_method}

Here we propose a generic solution template for the abstract problem defined above. The policy of each agent is composed of three modules: observation encoder, communicator and action selector. Below we describe each of these modules for an arbitrary agent $i$ in detail. While each agent has its own copy of these modules, we suppress the dependence of modules on agent index to avoid clutter. An instantiation of this template for the traffic management problem has been presented in Section \ref{section:intelligent_traffic_management}.


\textbf{Observation Encoder:}
\label{section:observation_encoder}
This module encodes the observation of an agent into a form suitable for the other two modules. We denote this module by $f_{\mathrm{obs}}$ and its output by $\mathbf{h}_i^{(t)}$, i.e., $\mathbf{h}_i^{(t)} = f_{\mathrm{obs}}(\mathbf{o}_i^{(t)})$.


\textbf{Communicator:}
\label{section:communicator}
The communicator module $f_{\mathrm{comm}}$ takes the encoded observation and a history of received messages as input and produces a message $\mathbf{m}_i^{(t)} \in \{0, 1\}^d$ to be broadcasted as output. Note that the messages are $d$-dimensional binary vectors. Aside from computing the message to be sent, $f_{\mathrm{comm}}$ also processes the messages received from the neighbors to generate a vector $\bar{\mathbf{q}}^{(t)} \in \mathbb{R}^d$ that summarizes the received messages.


\textbf{Action Selector:}
\label{section:action_selector}
The action selector module $f_{\mathrm{act}}$ takes the output of observation encoder $\mathbf{h}_i^{(t)}$ and an encoding of received messages $\bar{\mathbf{q}}_i^{(t)}$ as input and produces a probability distribution over actions in $\mathcal{A}_i$ as output. The action $\mathbf{a}_i^{(t)}$ is then sampled from this distribution.

These three modules jointly formulate the policy $\pi_i$ for agent $i$ that takes the current observation and messages from all neighbors as input and produces an action along with a message to be broadcasted to the neighbors as output. The policies are then optimized to maximize the expected rewards as described above.


\begin{figure}
	\centering
	\includegraphics[width=\linewidth]{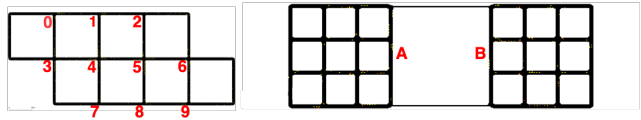}
	\caption{Traffic networks used in our experiments. On the left is a $10$ agent network (network 1) and on the right is a $28$ agent network (network 2) that has $14$ agents in each of the sub-networks (A \& B) connected by single 1-way lanes.}
	\label{fig:traffic_networks}
\end{figure}

\section{Intelligent Traffic Management}
\label{section:intelligent_traffic_management}

In this section, we first cast the problem of intelligently managing traffic controllers in the general framework presented in Section \ref{section:problem_setup} and then instantiate the solution template from Section \ref{section:proposed_method} to solve this problem. As noted earlier, in this context, the traffic junctions correspond to agents and these agents are connected to each other via roads which form the underlying network.

\textbf{Simulator:} We used a traffic simulator known as Simulation of Urban MObility (SUMO) \citep{SUMO} to simulate the traffic flow. The two road networks with which we experiment are given in Fig.~\ref{fig:traffic_networks}. All roads (except the connectors in network 2) are two lane roads. While the smaller $10$ agent network (which we call network $1$) allows us to easily study aspects like grounding of emergent language, the bigger $28$ agent network (network $2$) has two distinct structural communities and it yields further insights into the relationship between network topology and emergent communication.

\textbf{Observations:} Each agent observes an image representation of the traffic junction obtained by cropping a square patch of size $140px$ centered at that agent from the simulation window. The observation patches of neighboring agents do not overlap and have a considerable amount of space between them which makes the problem challenging due to severe partial observability. Within the observation patch, a queue length of at most 8 vehicles (4 per lane) can be observed. We do not explicitly provide any additional information like queue length on different lanes to the agents as a powerful enough observation encoder can in principle extract such information from the raw images.

\begin{figure}
	\centering
	\includegraphics[scale=0.35]{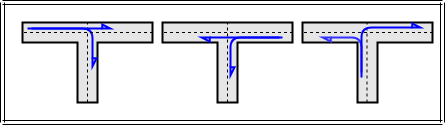}
	\includegraphics[scale=0.18]{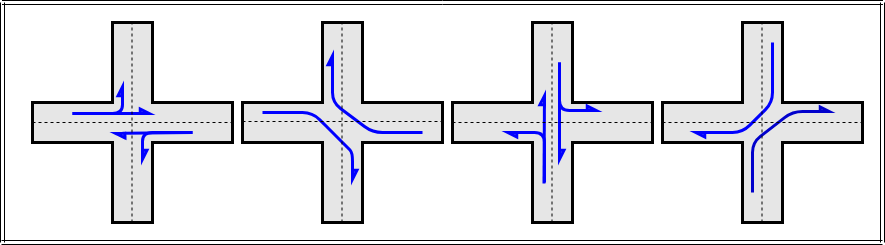}
	\caption{Permitted actions for 4-way and 3-way junctions. [Left-hand traffic]}
	\label{fig:traffic_actions}
\end{figure}

\textbf{Actions:} Fig.~\ref{fig:traffic_actions} shows the action space for agents residing on $4$-way and $3$-way junctions. Each allowed action corresponds to a particular configuration of traffic lights at the junction (also called a phase). Although the number of traffic light combinations is much higher, all other configurations are either not legal, or are unsafe.

\textbf{Rewards:} A vehicle is considered to be waiting at time $t$ if it is moving with a speed $<0.1m/s$. The reward for agent $i$ at time $t$ is computed as: 
\begin{equation}
	\label{eq:reward_function}
	r_i^{(t)} = -\Big(\ell_i^{(t)} + w_i^{(t)} - \sum_j d_{ij}^{(t)} + e_i^{(t)}\Big)
\end{equation}
where, at the given junction, $\ell_i^{(t)}$ is the number of waiting vehicles, $w_i^{(t)}$ is the sum of waiting times of all the waiting vehicles, $d_{ij}^{(t)}$ is the delay calculated as the ratio of the average speed of vehicles in lane $j$ and the maximum permitted speed and, $e_i^{(t)}$ is the number of times emergency braking was used. In our setting, we consider a linear combination of all rewards as they roughly have the same scale.

\textbf{Observation Encoder:} We model $f_{\mathrm{obs}}$ as a three-layered convolution neural network (CNN) \citep{Lecun98gradient-basedlearning}. As it is unreasonable to change the configuration of traffic lights very frequently, the actions are only taken once every five seconds. However, to obtain sufficient information about the environment, the observation is recorded at each second and the CNN output is aggregated using a Long Short-Term Memory (LSTM) network \citep{Hochreiter:1997:LSM:1246443.1246450} with hidden-size $64$. The output of LSTM after five time-steps is taken as a summary of agent's observation.

\textbf{Communicator:} Agents broadcast a message every second. Output message $\mathbf{m}_i^{(t)}$ is generated by passing the output of LSTM from observation encoder at time $t$ to a fully connected layer followed by application of Gumbel-Softmax \citep{journals/corr/JangGP16}. To aggregate received messages we use soft-attention. Let $\mathcal{U}_i$ be the set of neighbors of node $i$ in the network. At time $t$, agent $i$ receives messages broadcasted by its neighbors at the $(t - 1)^{th}$ time-step, i.e., $\mathbf{m}_j^{(t-1)}$ for $j \in \mathcal{U}_i$. The aggregate message encoding $\bar{\mathbf{q}}_i^{(t)}$ is generated as follows:
\begin{align}
    \mathbf{q}_i^{(t)} &= W \mathbf{h}_i^{(t)} \notag \\
    \bm{\alpha}_{i}^{(t)} &= \mathrm{softmax}\Big[{\mathbf{q}_{i}^{(t)}}^\intercal \mathrm{m}_{j}^{(t-1)} \,\,:\,\, j \in \mathcal{U}_i \Big] \notag \\
    \bar{\mathbf{q}}_i^{(t)} &= \sum_{j \in \mathcal{U}_i} \alpha_{ij}^{(t)} \mathrm{m}_j^{(t-1)}
    \label{eq:attention}
\end{align}
Here $W$ is a learnable parameter. This attention mechanism is in the same spirit as the query based attention used in \citep{DasEtAl:2019:TarMACTargetedMultiAgentCommunication}. The received message summaries, $\bar{\mathbf{q}}_i^{(t)}$, are then aggregated using a LSTM to provide a message history. This LSTM takes $\bar{\mathbf{q}}_i^{(t)}$ as input and produces $\hat{\mathbf{q}}_i^{(t)}$ as output. At every fifth time-step, $\hat{\mathbf{q}}_i^{(t)}$ is passed on to the action selector.

\textbf{Action Selector:} $f_{\mathrm{act}}$ takes the outputs of observation encoder and communicator as input. These vectors are passed through separate linear layers and the results are added to obtain a single vector of size $\vert\mathcal{A}_i\vert$. Softmax is applied on this vector to obtain a probability distribution over actions.

\textbf{Training details:} In all the modules, we use ReLU activation function after all linear layers unless it is followed by other activation functions like Softmax or Gumbel-Softmax. As actions are taken every five seconds, the rewards given in \eqref{eq:reward_function} are also accumulated over this period. Thus, if an action is taken at time $t$, the corresponding reward for this time-step is given by $\sum_{t^{'}=t}^{t + 4} r_i^{(t^{'})}$. We use the mean episode reward as the baseline and use the well known REINFORCE trick \citep{Williams:1992:SimpleStatisticalGradientFollowingAlgorithmsForConnectionistReinforcementLearning} to compute gradients for running the policy gradients algorithm \citep{SuttonEtAl:2000:PolicyGradientMethodsForReinforcementLearningWithFunctionApproximation}. We use Adam optimizer \citep{KingmaBa:2014:AdamAMethodforStochasticOptimization} with a learning rate of $10^{-4}$.


\section{Experiments}
\label{section:experiments}

We experimentally establish the following claims: \textbf{(i)} our approach outperforms baseline methods; \textbf{(ii)} agents exchange meaningful information; \textbf{(iii)} emergent communication is grounded in the actions taken by the agents; and \textbf{(iv)} network topology affects the nature of emergent communication.

\textbf{Comparison with baselines:} We compare our approach with the following baselines:

\textbf{(i)} Fixed-time control: The agents periodically switch between actions in a round-robin fashion after every five steps. This is how the presently deployed traffic controllers work.  

\textbf{(ii)} Self-Organizing Traffic Light control (SOTL): SOTL~\citep{SOTL} switches between actions when the queue length at an adjoining lane exceeds a predefined threshold (fixed to five in our implementation). This simple heuristic improves the performance of fixed time control.

\textbf{(iii)} Deep-Q Learning (DQN): Agents are training independently and each agent has its own deep Q-network. This baseline justifies the need of a multi-agent setup. 

\begin{figure}
    \begin{center}
        \includegraphics[width=0.40\textwidth]{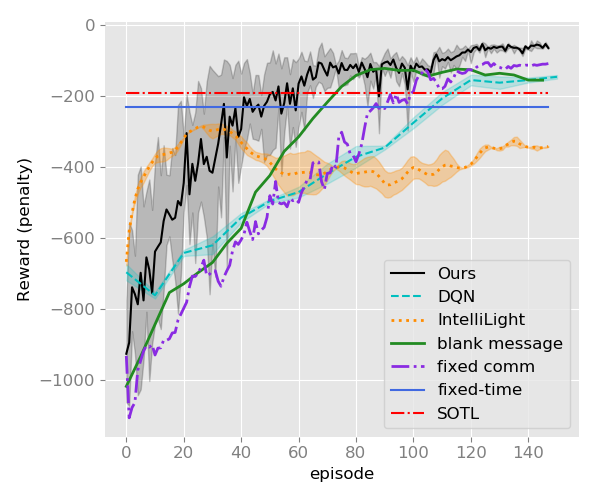}
    \end{center}
    \caption{Comparing our method with the baselines (Section \ref{section:experiments}). Five independent trained models were executed to collect episode rewards averaged over agents. The mean and standard deviation over five runs have been reported.}  
    \label{fig:convergence}
\end{figure}

\textbf{(iv)} IntelliLight: IntelliLight~\citep{Wei:2018} uses more elaborate observations that includes queue length, number of vehicles and updated waiting time at the adjoining lanes of a junction in addition to its image representation. As the action space used in \citep{Wei:2018} can lead to dangerous configurations, we used the action space specified in Section \ref{section:intelligent_traffic_management} in our implementation of this baseline.

\textbf{(v)} Fixed communication protocol: We modified the setup presented in Section \ref{section:intelligent_traffic_management} to use a fixed communication protocol. Agents do not learn to communicate as part of the training process, but rather share all the parameters needed to compute rewards in \eqref{eq:reward_function} with their neighbors directly. This baseline shows the utility of emergent communication.

Fig.~\ref{fig:convergence} compares our approach with these baselines. It can be seen that our method outperforms all baseline approaches. Also, fixed communication is better than no communication but it is not as good as the emergent communication.

\textbf{Robustness:}
In practice, roads are often blocked due to random events which changes the distribution of traffic. To study the robustness of our approach to such perturbations, we evaluated two types of trained models on test episodes, in which a randomly chosen road was blocked at the beginning of each episode (in network 1): \textbf{(i)} a model trained on a fixed network as described in Section \ref{section:intelligent_traffic_management} and, \textbf{(ii)} a model that was simultaneously trained on $25$ perturbed variants of the same road network. In case (ii), one of the $25$ perturbations was randomly sampled for each training episode. While (i) resulted in a reward of $\approx \mathbf{-177}$ points, (ii) performed considerably better, yielding a reward of $\approx \mathbf{-55}$ points. So, the proposed model can be made more robust by following the training procedure outlined in (ii). More interestingly, (ii) also has a regularization effect which improves the performance of the model even when network is fixed during testing.

\begin{table}[t]
    \centering
    \begin{tabular}{|c|c|c|c|}
        \hline
        \textbf{No. of Blind Agents} & \textbf{0} & \textbf{1} & \textbf{2} \\
        \hline
        \textbf{Reward} & -65 & -75 & -170 \\
        \hline
    \end{tabular}
    \caption{Effect of having different number of blind agents on average reward post convergence. Having one blind agent does not reduce the performance but having two blind agents does reduce it. Hence, communication is meaningful.}
    \label{table:blind}
\end{table}

\textbf{Utility of communication:}
We provide a qualitative analysis of the communication while following the guidelines presented in \citep{10.5555/3306127.3331757}. While the DQN and fixed communication protocol baselines indicate that the emergent communication is important, to further strengthen this argument, we performed two additional experiments: 

\textbf{(i)} We modified the setup presented in Section \ref{section:intelligent_traffic_management} to mask all communication messages in the system with an all zeros vector (\textit{blank message}. While this may seem very similar to the DQN baseline, we wanted to ensure that the improved performance of our method is not because of the use of a different training algorithm (Q-learning vs policy gradient) but because of communication. We not only noticed a drop in convergence speed but also observed that post convergence rewards were lower as compared to the original setting (the difference was $\approx\mathbf{85}$, also see Fig. \ref{fig:convergence}). 

\textbf{(ii)} We define an agent to be visually impaired (or blind) if it does not use its local observation while taking an action. Note that a visually impaired agent can still receive messages from its neighbors. We experimented with a setting where agent $4$ in Fig.~\ref{fig:traffic_networks} (network 1) was made visually impaired. We observed that, after convergence, the rewards were same as the rewards obtained in the original setup. This indicates that the visually impaired agent learned to receive necessary information from its neighbors through communication. To test this hypothesis further, we additionally made agent $5$ (which is a neighbor of agent $4$) visually impaired as well. As expected, since neighboring agents have been made visually impaired, they can no longer supplement each other's missing information using communication and hence the performance decreased. Table \ref{table:blind} summarizes these results.

\textbf{Grounding in communication:}
One reason for preferring the usage of discrete symbols for communication over continuous vectors is because we believe that it would be easier for humans to interpret discrete communication. This is because, in many cases, it may be possible to establish a relationship between words in the emergent language and physical entities/actions. An emergent language is said to be grounded if it satisfies this property. In many real world applications, like the intelligent traffic management problem, being able to interpret the process used for decision making is highly desirable. Our analysis shows that the emergent language in the proposed setup is grounded.

To establish this, we constructed a Pointwise Mutual Information (PMI)~\citep{Church:1990:WAN:89086.89095} matrix for each pair of agents. The rows of this matrix correspond to the actions of one agent (say $i$) and the columns correspond to the message sent by the other agent (say $j$). Let $\mathbf{P}^{ij} \in \mathbb{R}^{\vert\mathcal{A}_i\vert \times 2^d}$ denote the PMI matrix for the agents $i$ and $j$ as described above. A high value of $\mathbf{P}^{ij}_{kl}$ indicates that $k^{th}$ action taken by $i^{th}$ agent bears a strong relationship with $l^{th}$ word spoken by $j^{th}$ agent. For each pair of agents $(i, j)$, we computed the Singular Value Decomposition (SVD) of the PMI matrix $\mathbf{P}^{ij} = \mathrm{U}^{ij}\mathrm{S}^{ij}\mathrm{{V^{ij}}^T}$, where $\mathrm{U}^{ij} \in \mathbb{R}^{\vert \mathcal{A}_i \vert \times k}$, $\mathrm{S}^{ij} \in \mathbb{R}^{k \times k}$ and $\mathrm{V}^{ij} \in \mathbb{R}^{2^d \times k}$ and $k \leq \vert \mathcal{A}_i \vert$ is a hyper-parameter (we use $k=2$). The rows of $\mathrm{V}$ can be interpreted as representations of the words in the emergent language in the context of actions taken by the receiving agent. Under this representation, words will be similar if they lead to similar actions being taken by the receiving agent. 

\begin{figure}
    \centering  
\includegraphics[width=\linewidth]{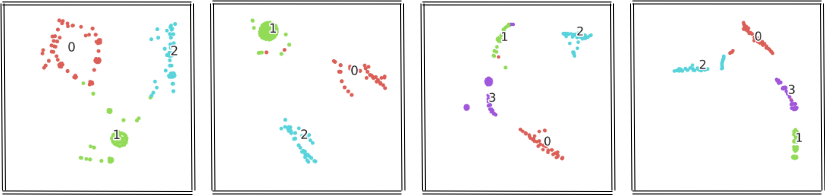}
\caption{\textbf{[Best viewed in colour]} Word embeddings of neighbour $j$ ($j \in \mathcal{U}_i$) corresponding to actions of agent $i$ for a 3-way junction (indexed by $\{0,1,2\}$) and for a 4-way junction (indexed by $\{0,1,2,3\}$. The clusters can be interpreted as distinct words used to mean different actions. 3-way junctions have an action space of size 3 and 4-way junctions have an action space of size 4.}
    \label{fig:word_action}
\end{figure}

Figure~\ref{fig:word_action} shows the t-SNE~\citep{7b54165e73a3424b8820136bcf61ca89} plot of the rows of $\mathrm{V}^{ij}$ matrix. Colors have been assigned to the points based on the action with which the word has the highest PMI. It can be seen that distinct clusters form, each corresponding to a different action. This signifies that neighboring agents use specific set of words to indicate actions, thus showing that the language is grounded.

\begin{figure}
    \centering \includegraphics[width=0.3\textwidth]{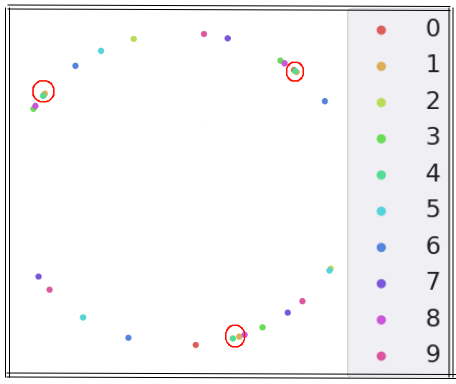}
    \caption{\textbf{[Best viewed in color]}. Action embeddings from matrix $\mathrm{U}$ (orthonormal matrix) of agent $0$ corresponding to messages from all agents. As highlighted by the red circles, the action embeddings of agent $0$ are overlapped for neighbors $(1,3)$. The color bar represents different agents.}
    \label{fig:action_word}
\end{figure}

Additionally, we plot the rows of $\mathrm{U}^{ij}$ matrix (with $k=2$), i.e., the action embeddings corresponding to the broadcasted messages. We set $i=0$ and pair agent $0$ with all agents $j=0, 1, \dots, 9$ to obtain different action embeddings for agent $0$ using $\mathbf{U}^{0j}$. In Fig.~\ref{fig:action_word}, we plot these action embeddings; different colors have been used for different agents $j=0, 1, 2, \dots, 9$. It can be observed that the points corresponding to neighbors $j=1, 3$ tend to be very close to each other (highlighted using red circles in Fig.~\ref{fig:action_word}). This implies that neighbors of agent $0$ are consistent in their use of messages for referring to actions. 

\begin{figure}[t]
\centering  
\includegraphics[width=\linewidth]{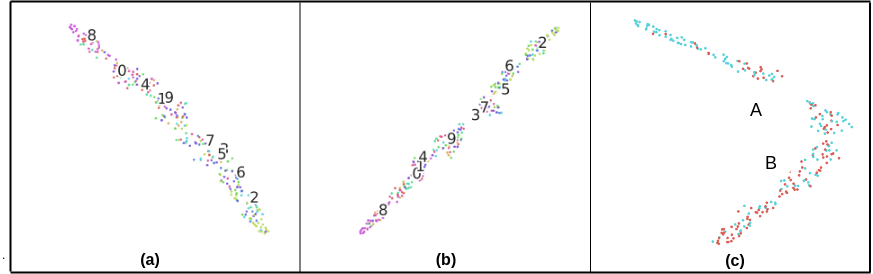}
\caption{\textbf{[Best viewed in color]}. Clustering of agents in the 10-agent network (network 1) for different runs [Fig.(a,b)]. The numbers denote the agents. Fig.(c) on the right represents the clustering in the 28-agent network (network 2) with A \& B denoting two 14-agent sub-networks. The position of the numbers denote the mean of the clusters.}
\label{fig:clusters}
\end{figure}

\textbf{Effect of network topology:} We obtain a tf-idf matrix where rows correspond to agents and columns correspond to the words in the vocabulary. On plotting a t-SNE plot of the rows of this matrix for agents in the ten node network (Fig.~\ref{fig:traffic_networks}), we noticed that agents that broadcast to the same neighbor tend to be clustered together. For instance, from Fig.~\ref{fig:clusters} [(a), (b)], one can infer that the following groups are formed:
\textbf{(i)} agents 0 \& 4 broadcasting to agent 1; \textbf{(ii)} agents 3 \& 7 broadcasting to agent 4; \textbf{(iii)} agents 3 \& 5 broadcasting to agent 4. It is also consistent with our findings in Fig.~\ref{fig:action_word}, where actions embeddings corresponding to the neighbors overlap.

\textbf{Experiments on larger networks:}
When agents are networked together, it is reasonable to expect that they will be influenced more by those agents with whom they are closely associated. While we have demonstrated that this holds for agents that are sending messages to a common receiver (Fig.~\ref{fig:clusters}), more broadly, one can understand the role played by communities in shaping the emergent communication.
 
To do so, we took a pair of networks, each having 14 agents and connect them by two single one-way lanea ($28$ agent network in Fig.~\ref{fig:traffic_networks}). As before, we plotted the t-SNE embeddings of the rows of tf-idf matrix (Fig.~\ref{fig:clusters} (c)) and observed that two clusters (denoted by A and B), corresponding to the two structural communities in the network are discovered. There is an overlapping region which we argue represents the common vocabulary used by agents from both communities.


\section{Conclusion}
\label{section:conclusion}

In this paper, we formulated a networked multi-agent reinforcement learning problem where cooperative agents communicate with each other using an emergent language. The problem of intelligent traffic management was cast in the general problem framework and empirical evaluations were made to: \textbf{(i)} demonstrate the utility of emergent communication in optimizing traffic flow; \textbf{(ii)} understand the properties of emergent language; and \textbf{(iii)} show the relationship between emergent communication and the underlying network topology. It would be interesting to extend this framework to dynamic graphs and we leave this for future work.


\bibliographystyle{ACM-Reference-Format}
\bibliography{main}

\end{document}